\documentclass[aps,prx,reprint,superscriptaddress,nofootinbib,longbibliography]{revtex4-2}

\usepackage{amsmath,amssymb,mathtools,bm}
\usepackage{graphicx}
\usepackage{dcolumn}
\usepackage{booktabs}
\usepackage{xcolor}
\usepackage[colorlinks=true,allcolors=blue!55!black]{hyperref}
\usepackage{microtype}
\newcommand{\dd}{\mathrm{d}}
\newcommand{\ii}{\mathrm{i}}

\newcommand{\NS}{\mathrm{NS}}
\newcommand{\R}{\mathrm{R}}

\newcommand{\eps}{\varepsilon}
\newcommand{\Aone}{\mathfrak{A}_{1}}
\newcommand{\Athree}{\mathfrak{A}_{3}}

\begin{document}

\title{An Effective String Theory Toolbox for Quantum Hall Interfaces II: \\ Majorana Fermions on Fluctuating Moore-Read Worldsheets}

\author{Ken K. W. Ma}
\affiliation{Independent researcher, Orlando, Florida 32837, USA}
\date{August 5, 2026}

\begin{abstract}
A Moore--Read interface carries a chiral Majorana mode on a boundary whose
geometry may itself fluctuate.  Fixed-edge theory does not determine how this
neutral mode should be transported when the interface bends and moves, or how
its dynamics couples to the fluctuating shape.  Here we construct a spatially
reparametrization-invariant Majorana theory on the nonrelativistic worldsheet
of a freely moving interface.  The changing line element fixes a universal
half-density transport law, while additional curvature- and
velocity-dependent couplings remain controlled by microscopic interface
physics.  The resulting framework identifies the Majorana stress tensor as
the mediator between neutral and geometric dynamics and provides the neutral
sector needed for effective theories of dynamical non-Abelian quantum Hall
interfaces.
\end{abstract}

\maketitle

\section{Introduction}
\label{sec:introduction}

The Moore--Read (Pfaffian) state supports a charged chiral boson and a neutral chiral
Majorana mode at its boundary~\cite{MooreRead1991,MilovanovicRead1996,ReadGreen2000}.
In the conventional edge problem, the boundary is fixed by a confining
potential, and both fields propagate on a prescribed line.  A freely moving
interface poses a different low-energy problem.  Its embedding is dynamical,
and normal motion changes the area occupied by the adjacent Hall fluids.  For
a self-bound interface this converts the charged sector into a geometric shape
mode with an asymptotically cubic dispersion~\cite{LiMa2021,TurkerYang2022,Ma2026I}.

Meanwhile, the neutral mode is not generated by the same deformation.  It descends from
the paired Ising sector of the Moore--Read phase and remains generically
linear at long wavelength, even when the charged branch is cubic.  Previous
interface theories could therefore treat the Majorana field as a decoupled
edge mode to leading order.  The remaining question is how that field is
transported when the line carrying it bends and moves, and which
geometry--fermion couplings are fixed by kinematics rather than by microscopic
interface physics.

In this work, we formulate the neutral sector on the nonrelativistic
worldsheet developed for the charged interface. Note that laboratory time remains physical, and only changes of the coordinate used to label points along the
curve are gauge redundancies.  Requiring a well-defined variational principle
on the time-dependent line element fixes the Majorana half-density transport
law and its canonical anticommutator.  The same construction separates the
kinematic transport term from curvature-dependent Hamiltonian couplings and
velocity-dependent mixed kinetic couplings.

The resulting theory gives two complementary limits.  In the leading
one-derivative, single-channel theory, a smooth static velocity profile can be
removed by a time-of-flight coordinate and a local field rescaling.  Static
curvature may shift propagation times and finite-size levels, but it cannot
reflect or bind a lone chiral Majorana as long as the local velocity remains
nonzero.  A time-dependent deformation instead couples to the Majorana stress
tensor and permits energy exchange with the shape mode.  At long wavelength,
the nonlocal part of this response is fixed by the Ising central charge
\(c_\psi=1/2\), although the overall geometric coupling is nonuniversal.

The single-Majorana description applies directly to an unreconstructed
Moore--Read--vacuum interface and, after charge localization, to the neutral
sector proposed for a Moore--Read--331 interface~\cite{Yang2017}.  More
complicated domain walls require a multichannel extension.  In particular,
Pfaffian--anti-Pfaffian interfaces can contain several neutral modes and a
nontrivial interfacial dipole~\cite{LevinHalperinRosenow2007,LeeRyuNayakFisher2007,MrossOregStern2018,ZhuShengYang2020,Hsin2020},
while microscopic studies find that reconstruction can be driven by the
fermionic sector and may remain confined to the Majorana sector near realistic
\(\nu=5/2\) edges~\cite{ZhangWuHutasoitJain2014,Lotric2025}.  We
therefore keep the one-channel results separate from claims that depend on a
specific reconstructed edge structure.

The manuscript is organized as follows.  Section~\ref{sec:setting} fixes the
worldsheet geometry and the topological data of the neutral sector.
Section~\ref{sec:covariant} derives the Majorana action and canonical
structure.  Section~\ref{sec:EFT} develops the geometric effective theory and
its consequences on prescribed curves.  Section~\ref{sec:shape_coupling}
couples the neutral mode to the dynamical shape field and derives the stress
response.  Multichannel interfaces and microscopic matching are discussed in
Sec.~\ref{sec:general_interfaces}. Technical derivations are collected in the Appendices.

\section{Neutral modes on a moving interface}
\label{sec:setting}

\subsection{Worldsheet geometry and mode content}

To start, we recall the geometry and gauge structure supplied by the companion Abelian theory~\cite{Ma2026I}.  A regular interface is described by the spatial embedding
\begin{equation}
 \bm X(\tau,\sigma)\in\mathbb R^2,
 \qquad
 \gamma=\bm X'^2,
 \qquad
 \dd s=\sqrt\gamma\,\dd\sigma,
 \label{eq:geometry_basic}
\end{equation}
where a prime denotes $\partial_\sigma$.  The unit tangent, outward normal,
and signed extrinsic curvature are
\begin{equation}
 \bm t=\partial_s\bm X,
 \qquad
 n^i=-\eps^{ij}t_j,
 \qquad
 K=\bm n\cdot\partial_s\bm t.
 \label{eq:tangent_normal_curvature}
\end{equation}
The velocity decomposes as
\begin{equation}
 \dot{\bm X}=v_t\bm t+v_n\bm n,
 \qquad
 \beta=\frac{v_t}{\sqrt\gamma},
 \qquad
 D_\tau=\partial_\tau-\beta\partial_\sigma.
 \label{eq:velocity_shift}
\end{equation}
The combination $D_\tau$ removes tangential relabeling and differentiates
along the physical motion of the curve.  The invariant measure evolves according to
\begin{equation}
 \partial_\tau\sqrt\gamma
 =\sqrt\gamma\left(\partial_sv_t-Kv_n\right).
 \label{eq:measure_evolution}
\end{equation}
Under an infinitesimal time-dependent relabeling
$\delta_\xi\bm X=\xi\bm X'$, a worldsheet scalar $f$ transforms as
$\delta_\xi f=\xi f'$, while $D_\tau f$ transforms covariantly. Thus, the preferred temporal and spatial derivatives are
\begin{equation}
 u^a\partial_a=D_\tau,
 \qquad
 s^a\partial_a=\partial_s.
 \label{eq:preferred_vectors}
\end{equation}
They encode the physical clock and the oriented spatial direction of the
nonrelativistic worldsheet; no Lorentz or Weyl symmetry is assumed.

The Moore--Read theory contains a charged \(U(1)\) sector and a neutral Ising
sector.  For an interface with vacuum, the charged sector is tied to the
physical motion of the boundary.  In the small-deformation regime the reduced bosonic
action has the form~\cite{LiMa2021,Ma2026I}
\begin{equation}
 \begin{split}
 S_b^{(2)}
 ={}&-\frac{B^2\Delta\nu}{4\pi}
 \int\dd t\,\dd x\,
 u\,\partial_x^{-1}\dot u
 \\
 &-\frac12\int\dd t\,\dd x\,
 \left[
 T_0u_x^2+2T_2u_{xx}^2+2T_{4b}u_{xxx}^2+\cdots
 \right],
 \end{split}
 \label{eq:bosonic_reduced_action}
\end{equation}
where $u(x,t)$ is the normal displacement and $\Delta\nu$ is the filling
fraction difference across the interface.  The corresponding bracket and
dispersion are
\begin{align}
 \{u(x),u(y)\}_D
 &=-\frac{2\pi}{B^2\Delta\nu}\partial_x\delta(x-y),
 \label{eq:bosonic_shape_bracket}
 \\
 \omega_b(k)
 &=\alpha_3k^3+\alpha_5k^5+\alpha_7k^7+\cdots,
 \label{eq:boson_dispersion_general}
\end{align}
with
\begin{equation}
 \alpha_3=\frac{2\pi T_0}{B^2\Delta\nu},
 \qquad
 \alpha_5=\frac{4\pi T_2}{B^2\Delta\nu}.
 \label{eq:alpha_coefficients}
\end{equation}

On the other hand, the neutral sector is described at long wavelength by a real chiral fermion $\psi=\psi^\dagger$ with central charge
\begin{equation}
 c_\psi=\frac12.
\end{equation}
On a straight interface its dispersion is generically
\begin{equation}
 \omega_\psi(k)
 =v_\psi k+\eta_3k^3+O(k^5).
 \label{eq:majorana_dispersion}
\end{equation}
The velocity $v_\psi$ and the higher-gradient coefficient $\eta_3$ are
nonuniversal.  Microscopic studies find that neutral and charged velocities
can differ strongly~\cite{WanYangRezayi2006,HuRezayiWanYang2009}, and the free-interface
calculations of Refs.~\cite{LiMa2021,TurkerYang2022} exhibit the same basic
separation between a linear neutral branch and a cubic shape branch.

The one-channel theory developed below applies to a Moore--Read--vacuum
interface when no additional neutral reconstruction occurs.  It also applies
to the Moore--Read--331 interface after charge localization leaves one chiral
Majorana mode~\cite{Yang2017}.  The shape dynamics of those two examples is
not identical. More specifically, $\Delta\nu\neq0$ for a Moore--Read--vacuum interface, whereas $\Delta\nu=0$ for Moore--Read and 331 states at the same filling.  In the
latter case electromagnetic Chern--Simons response does not by itself provide
a shape symplectic form.  The Majorana theory below remains valid on a moving
curve, but the dynamics of that curve must come from additional microscopic
or geometric-response data~\cite{ReadRezayi2011}.

\subsection{Anomaly, frame holonomy, and spin structure}
\label{sec:spin}

The net neutral chirality is fixed by the difference of chiral central charges
across the interface.  A right-moving
Majorana contributes $+1/2$ and a left-moving Majorana $-1/2$.  The thermal
Hall conductance is
\begin{equation}
 \frac{\kappa_{xy}}{T}
 =\frac{\pi^2k_B^2}{3h}\,c_-,
 \label{eq:thermal_Hall}
\end{equation}
where $c_-$ includes charged and neutral sectors~\cite{KaneFisher1997,CappelliHuertaZemba2002}.
The nonrelativistic worldsheet used below has a preferred laboratory time and
does not require a two-dimensional Lorentzian metric.  It is nevertheless
useful to recall the standard relativistic anomaly statement that identifies
the same chiral central charge.  If the neutral edge theory is coupled to an
auxiliary \(1+1\)-dimensional background metric, the covariant stress tensor
appropriate to bulk anomaly inflow obeys
\begin{equation}
 \nabla_aT^a{}_{b}
 =\frac{c_\psi}{96\pi}
 \eps_{bc}\nabla^c\mathcal R,
 \label{eq:gravitational_anomaly}
\end{equation}
up to sign conventions for chirality, curvature, and \(\eps_{ab}\)
~\cite{AlvarezGaumeWitten1984,Stone2012,BradlynRead2015}.  The consistent
stress tensor differs by the corresponding Bardeen--Zumino term.
Eq.~\eqref{eq:gravitational_anomaly} is not used to define the
nonrelativistic moving-worldsheet geometry.  Its role here is to identify the
anomaly coefficient.  In the complete Hall system the anomalous edge
variation is canceled by the difference of bulk gravitational responses.
The anomaly fixes the net chiral central charge, but not the propagation
velocity or any of the extrinsic-curvature coefficients introduced below.
Geometric response can nevertheless generate additional boundary terms, and
extrinsic geometry may affect which boundary interactions are allowed
~\cite{GromovJensenAbanov2016,SinhaBradlyn2025}.

We write the tangent vector as
\begin{equation}
 \bm t=(\cos\theta,\sin\theta).
 \label{eq:tangent_angle}
\end{equation}
The connection of the tangent--normal frame is
\begin{equation}
 \omega_\sigma=\theta'=\sqrt\gamma K,
 \qquad
 \omega_\tau=\dot\theta=\partial_sv_n+Kv_t.
 \label{eq:frame_connection}
\end{equation}
The angle \(\theta\) is measured relative to the chosen ambient frame,
while the adapted tangent--normal frame is fixed by the embedded curve
once its orientation is chosen. In the
\(p+\ii p\) description, a local redefinition of an auxiliary frame can be
compensated by the pairing phase.  This does not remove the geometric scalar
\(K\) of the embedded curve, and it does not fix the coefficient of an
independent local operator \(K T_\psi\).  The global frame holonomy is
nevertheless nontrivial:
\begin{equation}
 \oint\dd\sigma\,\omega_\sigma
 =\oint\dd s\,K
 =2\pi w,
 \label{eq:frame_holonomy}
\end{equation}
where $w$ is the turning number.  In the \(p+\ii p\) representative, the physical boundary condition depends on
the frame winding together with the pairing-phase winding and the bulk Ising
charge; it is not fixed by extrinsic curvature alone
~\cite{ReadGreen2000,StoneRoy2004,MorozHoyosRadzihovsky2016,Quelle2016,NayakReview2008}.

For a Moore--Read droplet, the neutral Hilbert space is organized by the Ising
primaries
\begin{equation}
 a\in\{1,\psi,\sigma\},
 \qquad
 h_1=0,
 \quad
 h_\psi=\frac12,
 \quad
 h_\sigma=\frac1{16}.
 \label{eq:Ising_weights}
\end{equation}
These conventions follow the standard Ising conformal field theory
normalization~\cite{DiFrancesco1997,Ginsparg1988}.
The $1$ and $\psi$ sectors use Neveu--Schwarz boundary conditions,
\begin{equation}
 \psi(s+L)=-\psi(s),
 \qquad a=1,\psi,
 \label{eq:NS_bc}
\end{equation}
whereas a bulk $\sigma$ quasiparticle changes the branch cut and gives the
Ramond sector
\begin{equation}
 \psi(s+L)=+\psi(s),
 \qquad a=\sigma.
 \label{eq:R_bc}
\end{equation}
This relation between bulk quasiparticles and edge spin structure follows
from the Moore--Read edge construction and from the $p+\ii p$ description of
the neutral sector~\cite{MilovanovicRead1996,WanYangRezayi2006,FendleyFisherNayak2007}.  The
Ramond sector contains a Majorana zero mode.  Its Hilbert-space interpretation
requires the global fermion-parity and fusion-space constraints of the full
topological theory; a single local edge field does not by itself define an
independent two-state system. The distinction needed below is
\begin{itemize}
\item local frame/phase convention: removable,
\item global spin structure: topological.
\end{itemize}
Extrinsic curvature may still enter the local Hamiltonian through
nonuniversal coefficients.  Those terms are classified in
Sec.~\ref{sec:EFT}.

\section{Majorana worldsheet action and canonical structure}
\label{sec:covariant}

The geometry in Sec.~\ref{sec:setting} fixes how a field is carried by the
moving interface.  The remaining task is to construct a first-order fermionic
action whose variational and canonical structures respect that transport.

\subsection{Geometric transport and quadratic action}

Let $\psi(\tau,\sigma)$ be a real Grassmann field that transforms as a scalar
under spatial relabeling.  The leading quadratic action on the moving curve is
\begin{equation}
 S_\psi^{(0)}
 =\frac{\ii}{2}
 \int\dd\tau\,\dd\sigma\,\sqrt\gamma\,
 \psi
 \left(
 D_\tau+v_\psi\partial_s-\eta_3\partial_s^3
 \right)
 \psi.
 \label{eq:majorana_covariant_action}
\end{equation}
On a closed interface, odd spatial derivatives are formally
anti-self-adjoint with respect to \(\int\dd s\).  The convective derivative
requires more care because the measure evolves with the curve.  Define its
formal adjoint with respect to the spacetime measure by
\begin{equation}
 \int\dd\tau\,\dd\sigma\,\sqrt\gamma\,fD_\tau g
 =\int\dd\tau\,\dd\sigma\,\sqrt\gamma\,
 (D_\tau^\dagger f)g,
 \label{eq:Dtau_adjoint_definition}
\end{equation}
up to temporal endpoint terms.  Integration by parts, together with
Eqs.~\eqref{eq:velocity_shift} and \eqref{eq:measure_evolution}, gives
\begin{equation}
 D_\tau^\dagger
 =-D_\tau+Kv_n.
 \label{eq:Dtau_adjoint}
\end{equation}
Consequently, variation of Eq.~\eqref{eq:majorana_covariant_action} gives
\begin{equation}
 \left[
 D_\tau-\frac12Kv_n
 +v_\psi\partial_s
 -\eta_3\partial_s^3
 \right]\psi=0.
 \label{eq:majorana_geometric_eom}
\end{equation}
The term \(-Kv_n/2\) is fixed by transport on the changing line element.  It is
not an adjustable curvature coupling: it preserves the invariant equal-time
pairing and, after quantization, the equal-time anticommutator.

For a position-dependent coefficient $f(s,\tau)$, define the anti-self-adjoint
operators
\begin{align}
 \Aone[f]
 &=f\partial_s+\frac12\partial_sf,
 \label{eq:A1_definition}
 \\
 \Athree[f]
 &=\frac12\left(f\partial_s^3+\partial_s^3f\right)
 \nonumber\\
 &=f\partial_s^3
 +\frac32f_s\partial_s^2
 +\frac32f_{ss}\partial_s
 +\frac12f_{sss}.
 \label{eq:A3_definition}
\end{align}
With position-dependent coefficients, the quadratic equation takes the form
\begin{equation}
 \left[
 D_\tau-\frac12Kv_n
 +\Aone[V]-\Athree[\eta_{\mathrm{eff}}]+\cdots
 \right]\psi=0,
 \label{eq:general_majorana_equation}
\end{equation}
where $V$ is the local propagation velocity and $\eta_{\mathrm{eff}}$ is the leading
dispersive coefficient.  The variation and adjoint identities are derived in
Appendix~\ref{app:variation}.

\subsection{Half-density variables and relabeling algebra}

Canonical normalization is obtained with the spatial half-density
\begin{equation}
 \chi=\gamma^{1/4}\psi.
 \label{eq:half_density_definition}
\end{equation}
Under $\sigma\mapsto\sigma+\xi$, it transforms as
\begin{equation}
 \delta_\xi\chi
 =\xi\chi'+\frac12\xi'\chi.
 \label{eq:half_density_transform}
\end{equation}
Define the Lie derivative of a half-density along the one-dimensional vector
$f(\sigma)\partial_\sigma$ by
\begin{equation}
 \mathcal L_f^{(1/2)}
 =f\partial_\sigma+\frac12f'.
 \label{eq:half_density_Lie}
\end{equation}
For the first-derivative theory with a local velocity \(V\), the scalar-field
action is equivalent to
\begin{equation}
 S_{\chi}^{(0)}
 =\frac{\ii}{2}
 \int\dd\tau\,\dd\sigma\,
 \chi
 \left[
 \partial_\tau
 -\mathcal L_\beta^{(1/2)}
 +\mathcal L_{V/\sqrt\gamma}^{(1/2)}
 \right]\chi.
 \label{eq:half_density_action}
\end{equation}
In this variable the kinetic term has a fixed measure, and the canonical
bracket is immediate.

The momentum conjugate to $\chi$ is
\begin{equation}
 \pi_\chi=\frac{\ii}{2}\chi,
\end{equation}
so the fermionic primary constraint is second class.  With the standard
graded canonical bracket, the reduced classical bracket is
\begin{equation}
 \{\chi(\sigma),\chi(\sigma')\}_D
 =-\ii\delta(\sigma-\sigma').
 \label{eq:chi_Dirac_bracket}
\end{equation}
Quantization gives
\begin{equation}
 \{\widehat\chi(\sigma),\widehat\chi(\sigma')\}_+
 =\delta(\sigma-\sigma').
 \label{eq:chi_quantum_bracket}
\end{equation}
Equivalently, in invariant variables,
\begin{equation}
 \{\widehat\psi(s),\widehat\psi(s')\}_+
 =\delta_s(s-s'),
 \label{eq:psi_invariant_bracket}
\end{equation}
where $\delta_s$ is the delta function defined with respect to $\dd s$.

The corresponding contribution to the spatial-reparametrization generator is
\begin{equation}
 G_\psi[\xi]
 =\frac{\ii}{2}
 \int\dd\sigma\,\xi\chi\chi'.
 \label{eq:fermion_diff_generator}
\end{equation}
Using Eq.~\eqref{eq:chi_Dirac_bracket},
\begin{equation}
 \{\chi,G_\psi[\xi]\}_D
 =\xi\chi'+\frac12\xi'\chi,
 \label{eq:fermion_diff_action}
\end{equation}
and the classical generators satisfy the Witt algebra
\begin{equation}
 \{G_\psi[\xi_1],G_\psi[\xi_2]\}_D
 =G_\psi[\xi_1\xi_2'-\xi_2\xi_1'].
 \label{eq:fermion_Witt}
\end{equation}
Adding this term to the embedding and charged-sector generators gives the
single classical spatial-reparametrization constraint of the full worldsheet.
The Majorana field is matter under that symmetry; it introduces no additional
gauge redundancy.  The graded Dirac reduction and the action of the
relabeling generator are detailed in Appendix~\ref{app:canonical}.

After choosing the spatial coordinate, the normal-ordered modes of the
physical neutral stress tensor obey the Virasoro algebra
\begin{equation}
 [L_m,L_n]
 =(m-n)L_{m+n}
 +\frac{c_\psi}{12}(m^3-m)\delta_{m+n,0},
 \label{eq:majorana_Virasoro}
\end{equation}
with \(c_\psi=1/2\)~\cite{DiFrancesco1997,Ginsparg1988}.  This operator
algebra should be distinguished from the classical first-class gauge
constraint above.  Incorporating it into a fully anomaly-free quantum
worldsheet constraint requires the bulk gravitational response and the
associated inflow, which are not constructed explicitly here.

\section{Geometric effective theory and prescribed interfaces}
\label{sec:EFT}

Once the transport law is fixed, geometry can enter the neutral Hamiltonian
through local operators.  We first organize those operators and then examine
what they imply when the interface shape is prescribed.

\subsection{Local operator basis and power counting}

Fermion parity restricts the local action to even powers of the Majorana
field.  At quadratic order, only the anti-self-adjoint part of a
differential operator contributes to
$\frac{\ii}{2}\int\psi\mathcal O\psi$.  Consequently, even spatial derivative
operators are not independent: after integration by parts they reduce to odd
operators with derivatives acting on their coefficients.  Through third
spatial derivative order, the general one-component action can be organized
as
\begin{equation}
 S_\psi
 =\frac{\ii}{2}
 \int\dd\tau\,\dd\sigma\sqrt\gamma\,
 \psi
 \left[
 D_\tau+\Aone[V]-\Athree[\eta_{\mathrm{eff}}]
 \right]\psi
 +O(\partial^5).
 \label{eq:general_EFT_action}
\end{equation}
A convenient off-shell derivative expansion is
\begin{equation}
 \begin{split}
 V={}&v_\psi
 +\lambda_KK
 +\lambda_{K^2}K^2
 +\lambda_{sK}\partial_sK
 \\
 &+\lambda_nv_n
 +\lambda_{Kn}Kv_n
 +\lambda_{n^2}v_n^2
 +\cdots.
 \end{split}
 \label{eq:V_expansion}
\end{equation}
The first line contains geometry-dependent Hamiltonian data on a prescribed
interface.  Once the shape is dynamical, the terms involving \(v_n\) are
mixed kinetic couplings because \(v_n\) contains the shape velocity.  They can
modify the canonical structure of the combined shape--Majorana system even
though the Majorana constraint itself is unchanged.  Meanwhile,
\begin{equation}
 \eta_{\mathrm{eff}}=\eta_3+O(K,v_n,\partial_sK).
 \label{eq:Eta_expansion}
\end{equation}
Terms in the omitted part of $\eta_{\mathrm{eff}}$ enter at higher combined derivative order.
Acceleration-dependent terms such as $D_\tau K$ may also be written, but at a
fixed order some can be moved into Eq.~\eqref{eq:V_expansion} by integration by
parts, field redefinitions, and the leading shape equation of motion.
 Mixed operators containing additional $D_\tau$ derivatives can likewise be
reduced order by order using the leading chiral equation, provided no extra
low-energy fermion branch is introduced.  Eq.~\eqref{eq:general_EFT_action} is an operator basis for the quadratic
one-channel theory.  Interactions with more fermion fields are discussed
below.

None of the coefficients in Eq.~\eqref{eq:V_expansion} is fixed by the Ising
topological order.  They depend on the microscopic interface profile and on
the convention used to define the sharp dividing curve.  Their allowed or
forbidden status can nevertheless be constrained by symmetries.  For an
oriented Moore--Read--vacuum interface there is no symmetry that exchanges the
two sides, so terms odd in $K$ are allowed.  For an interface with a
microscopic side-exchange symmetry, the restriction depends on how that
symmetry acts on the oriented tangent, the Majorana field, and the chiral
direction.  If it sends \(K\mapsto-K\) and \(v_n\mapsto-v_n\) while
leaving the oriented stress operator invariant, the corresponding
odd-curvature and odd-velocity coefficients vanish.  Other implementations
that also reverse \(\partial_s\) or the chirality must be analyzed
separately.

At the straight Majorana fixed point, the fermion has scaling dimension
$[\psi]=1/2$ and the stress density
\begin{equation}
 T_\psi
 =-\frac{\ii}{2}\psi\partial_s\psi
 \label{eq:stress_density_definition}
\end{equation}
has dimension two.  Since $K$ has one spatial derivative, the leading
interaction
\begin{equation}
 H_{K\psi\psi}
 =\lambda_K\int\dd s\,K T_\psi
 \label{eq:leading_KT_interaction}
\end{equation}
has operator dimension three when the geometry is treated as a source for the
neutral \(z=1\) fixed point, and is therefore irrelevant in that source
counting.  The full dynamical problem contains both a \(z=1\) Majorana branch
and a \(z=3\) shape branch, so the mixed theory does not admit a single
isotropic scaling assignment.  Its asymptotic decoupling is established more
directly below by the energy mismatch and the explicit stress response.
The coupling remains observable: it controls finite-size shifts, stress-tensor spectroscopy, and
the leading neutral-loop correction to the shape dynamics.

The geometric transport term $-Kv_n/2$ in
Eq.~\eqref{eq:majorana_geometric_eom} should not be confused with
$\lambda_K K T_\psi$.  The former is fixed kinematically and preserves the
invariant equal-time pairing on a changing line element.  The latter changes
the local
propagation velocity and is a nonuniversal Hamiltonian coupling.  A term
\(v_nT_\psi\), by contrast, is a nonuniversal mixed kinetic coupling when the
shape is dynamical.

Fermion parity also permits higher-dimensional Ising descendants, including
operators represented schematically by \(:T_\psi^2:\).  In a free chiral
Majorana theory their independent content must be reduced using total
derivatives, free-fermion identities, and null-state relations.  Such
operators have higher scaling dimension than the quadratic curvature
coupling and do not alter the universal long-wavelength results derived
below.  For several Majoranas, flavor-current
interactions can be marginal or relevant when counterpropagating sectors are
present; those effects belong to the multichannel theory of
Sec.~\ref{sec:general_interfaces}.

\subsection{Static one-channel propagation and finite-size spectrum}
\label{sec:propagation}

Consider the one-derivative theory on a static curve, \(v_n=v_t=0\), with a
smooth positive local velocity \(V(s)\).  Eq.~\eqref{eq:general_majorana_equation} becomes
\begin{equation}
 \left[
 \partial_t+V(s)\partial_s+\frac12V_s(s)
 \right]\psi=0.
 \label{eq:static_variable_velocity_eom}
\end{equation}
Define the time-of-flight coordinate and rescaled field
\begin{equation}
 y(s)=\int^s\frac{\dd s'}{V(s')},
 \qquad
 \widetilde\psi(y,t)=\sqrt{V(s)}\,\psi(s,t).
 \label{eq:time_of_flight_transform}
\end{equation}
Then
\begin{equation}
 \left(\partial_t+\partial_y\right)\widetilde\psi=0.
 \label{eq:uniform_chiral_eom}
\end{equation}
Thus every smooth static profile that enters only through \(V(s)\) is exactly
equivalent to uniform chiral propagation.  It changes the local normalization
and the accumulated time of flight, but it produces neither reflection nor
localization as long as \(V(s)\) remains finite and does not change sign.

Bound states therefore require ingredients absent from this truncation, such
as a counterpropagating partner and a mass domain wall, an endpoint or vortex,
a zero of \(V(s)\), or higher-derivative terms that generate additional
low-energy roots.

For a closed loop, define the total time of flight
\begin{equation}
 \mathcal T=\oint\frac{\dd s}{V(s)}.
 \label{eq:total_time_of_flight}
\end{equation}
The spectrum of the first-derivative theory is exactly
\begin{equation}
 \omega_n=\frac{2\pi}{\mathcal T}(n+\alpha),
 \qquad n\in\mathbb Z,
 \qquad
 \alpha=
 \begin{cases}
 1/2,&\NS,\\
 0,&\R,
 \end{cases}
 \label{eq:tof_spectrum}
\end{equation}
with the topological projection discussed in Sec.~\ref{sec:spin}.

For a circle of radius $R$ and signed curvature $K_R=s_K/R$, the local
velocity is constant:
\begin{equation}
 V_R
 =v_\psi
 +\frac{s_K\lambda_K}{R}
 +\frac{\lambda_{K^2}}{R^2}
 +O(R^{-3}).
 \label{eq:circle_velocity}
\end{equation}
The Majorana modes are labeled by $r\in\mathbb Z+1/2$ in the NS sector and
$r\in\mathbb Z$ in the R sector.  Including the leading cubic dispersion,
\begin{equation}
 \epsilon_r(R)
 =\frac{V_R}{R}r
 +\frac{\eta_3}{R^3}r^3
 +O(R^{-4}).
 \label{eq:circle_single_particle_spectrum}
\end{equation}
Equivalently,
\begin{equation}
 \epsilon_r(R)
 =\frac{v_\psi}{R}r
 +\frac{s_K\lambda_K}{R^2}r
 +\frac{\lambda_{K^2}r+\eta_3r^3}{R^3}
 +\cdots.
 \label{eq:circle_spectrum_expanded}
\end{equation}
The radius dependence separates the leading geometric coefficients in a
finite-size spectrum.

For a state in topological sector \(a\), the conformal contribution to the
many-body energy is
\begin{equation}
 E_{a,N}^{\rm neutral,conf}(R)
 =\frac{V_R}{R}
 \left(h_a+N-\frac{c_\psi}{24}\right).
 \label{eq:circle_CFT_spectrum}
\end{equation}
where $N$ is a nonnegative descendant level.  The conformal weights and the
Casimir term are universal, while \(V_R\) is not
~\cite{DiFrancesco1997,Ginsparg1988}.  The displayed factor \(V_R/R\)
already includes the curvature contribution proportional to
\(\lambda_{K^2}\).  The leading nonconformal corrections enter at order
\(R^{-3}\); in particular, \(\eta_3\) produces state-dependent splittings that
are not fixed by \(h_a+N\) alone.  Our mode and normal-ordering
conventions on the circle are collected in Appendix~\ref{app:circle}.  The R
sector contains the zero mode $r=0$, whose parity and degeneracy are fixed only
after the bulk topological charge and the charged sector are included.

\section{Coupling to the fluctuating shape}
\label{sec:shape_coupling}

The preceding section treated geometry as an external background.  We now let
the interface fluctuate and determine the leading interaction between the
cubic shape mode and the neutral Majorana sector.

\subsection{Leading shape--Majorana interaction}

Take the local static gauge
\begin{equation}
 \bm X(t,x)=(x,u(t,x)).
 \label{eq:Monge_embedding}
\end{equation}
Here subscripts denote derivatives with respect to \(x\), while a dot denotes
\(\partial_t\). The exact geometric quantities are
\begin{align}
 \sqrt\gamma&=\sqrt{1+u_x^2},
 \\
 K&=\frac{u_{xx}}{(1+u_x^2)^{3/2}},
 \\
 v_n&=\frac{\dot u}{\sqrt{1+u_x^2}},
 \\
 \beta&=\frac{u_x\dot u}{1+u_x^2}.
 \label{eq:Monge_geometric_data}
\end{align}
For constant $v_\psi$ and no explicit geometric coefficient, the first-derivative Majorana action becomes
\begin{equation}
 \begin{split}
 S_{\psi,\mathrm{kin}}
 =\frac{\ii}{2}\int\dd t\,\dd x
 \Bigl[&
 \sqrt{1+u_x^2}\,\psi\dot\psi
 \\
 &-\frac{u_x\dot u}{\sqrt{1+u_x^2}}\,\psi\psi_x
 +v_\psi\psi\psi_x
 \Bigr].
 \end{split}
 \label{eq:exact_Monge_kinematic_action}
\end{equation}
Pure worldsheet kinematics therefore generates no interaction linear in the
shape displacement.  Expanding Eq.~\eqref{eq:exact_Monge_kinematic_action},
\begin{equation}
 \begin{split}
 S_{\psi,\mathrm{kin}}
 =S_{\psi,0}
 &+\frac{\ii}{2}\int\dd t\,\dd x
 \Bigl[
 \frac12u_x^2\psi\dot\psi
    -u_x\dot u\,\psi\psi_x
 \Bigr]
 \\
 &+O(u^4).
 \end{split}
 \label{eq:kinematic_quadratic_vertices}
\end{equation}
The leading linear interaction must instead come from a nonuniversal geometric
coefficient such as \(\lambda_K\).

Using Eq.~\eqref{eq:V_expansion}, the terms linear in $u$ are
\begin{equation}
 S_{u\psi\psi}^{(1)}
 =\frac{\ii}{2}
 \int\dd t\,\dd x\,
 \left(
 \lambda_Ku_{xx}
 +\lambda_n\dot u
 +\lambda_{sK}u_{xxx}
 \right)
 \psi\psi_x.
 \label{eq:linear_shape_majorana_vertex}
\end{equation}
The Monge-gauge expansion is verified in Appendix~\ref{app:Monge}.  The
\(\lambda_n\dot u\,T_\psi\) term is a mixed kinetic coupling and shifts the
shape momentum by an operator proportional to \(T_\psi\); it should not be
interpreted as an ordinary potential-energy term.
In terms of the stress density in Eq.~\eqref{eq:stress_density_definition},
\begin{equation}
 S_{u\psi\psi}^{(1)}
 =-\int\dd t\,\dd x\,
 \left(
 \lambda_Ku_{xx}
 +\lambda_n\dot u
 +\lambda_{sK}u_{xxx}
 \right)T_\psi.
 \label{eq:linear_vertex_stress_form}
\end{equation}
On the leading bosonic equation of motion,
$\dot u=\alpha_3u_{xxx}+\cdots$, the $\lambda_n$ and $\lambda_{sK}$ terms
enter on-shell observables through the combination
$\lambda_{sK}+\alpha_3\lambda_n$ at this order.  Off shell, for an externally
driven interface, they are distinct response coefficients.

\subsection{Low-energy stability and stress response}
\label{sec:response}

The curvature vertex permits a finite-momentum shape quantum of momentum
\(k\) to couple to two Majorana excitations with momenta \(q\) and
\(k-q\).  Although the shape branch descends from the charged \(U(1)\)
sector, its nonzero-momentum oscillator quanta do not change the total
electric charge; the charge zero mode is held fixed.  The process considered
below therefore respects charge conservation, while the bilinear Majorana
vertex also preserves fermion parity.  For a co-propagating right-moving
branch and \(0<q<k\),
\begin{equation}
 \begin{split}
 &\omega_\psi(q)+\omega_\psi(k-q)
 \\
 =~&v_\psi k
 +\eta_3\left[q^3+(k-q)^3\right]
 +\cdots.
 \end{split}
 \label{eq:two_majorana_energy}
\end{equation}
The bosonic energy is $\omega_b(k)=\alpha_3k^3+O(k^5)$.  Therefore, for
$v_\psi\neq0$,
\begin{equation}
 \omega_b(k)\ll
 \omega_\psi(q)+\omega_\psi(k-q)
 \qquad(k\rightarrow0).
 \label{eq:decay_forbidden}
\end{equation}
A long-wavelength shape quantum cannot decay into two co-propagating linearly
dispersing Majoranas.  The decay is therefore kinematically forbidden in the asymptotic
long-wavelength regime.  If the signs of the dispersions permit a crossing,
its characteristic momentum is of order
\begin{equation}
 k_*\sim\sqrt{\left|\frac{v_\psi}{\alpha_3}\right|},
 \label{eq:crossing_scale}
\end{equation}
which may lie outside the sharp-interface effective-theory regime.  Fermionic
curvature, additional branches, or a tuned small $v_\psi$ can modify this
conclusion at finite momentum.

A time-dependent external shape deformation is different.  It supplies
energy independently of the bosonic on-shell relation and can resonantly
create neutral excitations.  The relevant response is determined by the
Majorana stress tensor.

For a straight interface with linear Majorana dispersion, define
\begin{equation}
 \chi^R_{TT}(t,x)
 =-\ii\Theta(t)
 \langle[T_\psi(t,x),T_\psi(0,0)]\rangle.
 \label{eq:TT_retarded_definition}
\end{equation}
The equal-time Virasoro commutator contains the central term
\begin{equation}
 [T_\psi(x),T_\psi(y)]
 =\cdots
 +\frac{\ii c_\psi}{24\pi}
 \partial_x^3\delta(x-y),
 \label{eq:stress_commutator}
\end{equation}
where the omitted terms are proportional to the stress tensor itself.  In the
vacuum on the infinite line, those terms have zero expectation value.  With
the Fourier convention
$e^{\ii\omega t-\ii kx}$, chiral propagation gives
\begin{equation}
 \chi^R_{TT}(\omega,k)
 =\frac{c_\psi}{24\pi}
 \frac{k^3}{\omega-v_\psi k+\ii0^+}
 +\chi_{\rm contact},
 \label{eq:TT_response_full}
\end{equation}
where \(\chi_{\rm contact}\) is a local, regulator-dependent polynomial.
The pole and its residue are fixed by chirality and \(c_\psi=1/2\).  The
stress correlator is derived in Appendix~\ref{app:stress}.  The corresponding
spectral density is
\begin{equation}
 -2\operatorname{Im}\chi^R_{TT}(\omega,k)
 =\frac{c_\psi}{12}k^3
 \delta(\omega-v_\psi k)
 \label{eq:TT_spectral_density}
\end{equation}
for a right-moving ideal linear branch.  A time-dependent curvature harmonic can therefore excite the neutral channel
at \(\omega=v_\psi k\).  Dispersion, disorder, or additional neutral modes
broaden or split this ideal resonance.

The leading interaction is \(-\lambda_K\int u_{xx}T_\psi\).  We define the
retarded induced kernel by
\(D_R^{-1}=D_{0,R}^{-1}+\Pi_u^R\).  To quadratic order in \(u\), integrating
out the Majorana mode then gives
\begin{equation}
 \Pi_u^R(\omega,k)
 =\lambda_K^2k^4\chi^R_{TT}(\omega,k).
 \label{eq:shape_self_energy_general}
\end{equation}
Using Eq.~\eqref{eq:TT_response_full},
\begin{equation}
 \begin{aligned}
 \Pi_u^R(\omega,k)
 ={}&\frac{c_\psi\lambda_K^2}{24\pi}
 \frac{k^7}{\omega-v_\psi k+\ii0^+}
 \\
 &+\text{local contact terms}.
 \end{aligned}
 \label{eq:shape_self_energy_boxed}
\end{equation}
This pole is the leading universal nonlocal contribution of the neutral
sector.  Evaluated on the long-wavelength bosonic pole,
$\omega_b=\alpha_3k^3+\cdots$,
\begin{equation}
 \Pi_u^R(\omega_b,k)
 =-\frac{c_\psi\lambda_K^2}{24\pi v_\psi}k^6
 +O(k^8).
 \label{eq:self_energy_low_k}
\end{equation}
There is no imaginary part at asymptotically small \(k\), in agreement with the
kinematic argument in Eq.~\eqref{eq:decay_forbidden}.  Expanding the nonlocal
pole on the long-wavelength bosonic branch produces an analytic \(k^6\)
contribution to the quadratic shape kernel and therefore contributes to the
\(k^7\) dispersion coefficient.  That analytic coefficient cannot be
separated invariantly from local contact terms, microscopic neutral-sector
renormalizations, or the bare \(T_{4b}\) term.  The unexpanded pole structure
and its central-charge-controlled residue in
Eq.~\eqref{eq:shape_self_energy_boxed} are physical.

For a general linear vertex in Eq.~\eqref{eq:linear_vertex_stress_form}, the
factor $\lambda_Kk^2$ in Eq.~\eqref{eq:shape_self_energy_general} is replaced
by the full source
\begin{equation}
 g(\omega,k)
 =\lambda_Kk^2+\ii\lambda_n\omega+\ii\lambda_{sK}k^3,
 \label{eq:general_stress_source}
\end{equation}
with $\Pi_u^R=g(-\omega,-k)g(\omega,k)\chi^R_{TT}$.

\subsection{Topological-sector spectroscopy on a droplet}

On a circular droplet, a nonuniform shape deformation couples curvature
harmonics to the Majorana stress tensor.  In the following transition
amplitudes, we quotient out the purely reparametrization-induced redistribution
of the line element and isolate the physical source generated by the
extrinsic-curvature dependence of \(V\).  For a normal displacement
$u(\theta)=u_n e^{\ii n\theta}$, the linear curvature variation is
\begin{equation}
 \delta K_n
 =\frac{1-n^2}{R^2}u_n e^{\ii n\theta}
 \label{eq:circle_curvature_form_factor}
\end{equation}
up to the orientation sign.  The factor $1-n^2$ removes the rigid-translation
mode $n=1$ before any CFT selection rule is applied.  Let $L_n$ denote the
Virasoro generators and let $|h_a\rangle$ be an Ising primary state.  For
$n>0$, the remaining harmonic drive contains a term proportional to
$L_{-n}$, and the exact CFT transition norm is
\begin{equation}
 \begin{split}
 \|L_{-n}|h_a\rangle\|^2
 &\equiv
 \langle h_a|L_nL_{-n}|h_a\rangle
 \\
 &=2nh_a
 +\frac{c_\psi}{12}n(n^2-1).
 \end{split}
 \label{eq:Virasoro_transition_norm}
\end{equation}
This follows directly from Eq.~\eqref{eq:majorana_Virasoro}; the derivation
is included in Appendix~\ref{app:stress}.  For the three Ising sectors,
\begin{equation}
 h_a=0,\ \frac12,\ \frac1{16},
 \qquad c_\psi=\frac12.
 \label{eq:Ising_spectroscopy_values}
\end{equation}
The curvature-driven neutral spectral weight is therefore topological-sector
dependent even though the coupling coefficient $\lambda_K$ is nonuniversal.
For the identity-sector vacuum, $L_{-1}|0\rangle=0$ and the first
nontrivial stress descendant occurs at $n=2$.  At that harmonic the three
CFT norms obey
\begin{equation}
 \mathcal W_{1,2}:\mathcal W_{\psi,2}:\mathcal W_{\sigma,2}
 =1:9:2,
 \label{eq:n2_Ising_weight_ratio}
\end{equation}
where, at leading order in the large-$R$ derivative expansion, the common
geometric factor $(n^2-1)^2|u_n|^2$ and the unknown $\lambda_K^2$ have been
divided out.  At finite $R$, $\lambda_K$ is replaced by an effective source
coefficient containing the higher-curvature and mixed kinetic couplings.  This
factor remains common to the three Ising sectors and therefore cancels in the
ratio.  The $n=1$ response vanishes in an actual shape drive both because it
is a translation and, in the identity module, because the level-one
descendant is null.

Eq.~\eqref{eq:n2_Ising_weight_ratio} provides a direct numerical diagnostic:
a weak quadrupolar boundary harmonic can probe the neutral absorption, or an
avoided crossing between a shape oscillator and the corresponding Majorana
stress descendant, in a fixed topological sector.  The quantities
\(\mathcal W_{a,2}\) determine the relative oscillator strengths of these
transitions.  Ratios between sectors eliminate the unknown common geometric
source and isolate the neutral Ising matrix elements.  If the response is
resolved through an avoided crossing, the squared gaps obey the ratio
\(1:9:2\), while the gaps themselves scale as \(1:3:\sqrt{2}\).  Their
realization in a physical Moore--Read droplet must respect the compatible
charged sector, electron-parity projection, and total bulk topological charge;
the three neutral modules are not independently selectable states of one
fixed microscopic droplet.

\section{Multichannel interfaces and microscopic matching}
\label{sec:general_interfaces}

\subsection{Beyond a single Majorana channel}

Many interfaces require several Majorana fields \(\psi^A\).  In a
canonically normalized Majorana basis, the most general quadratic
first-derivative action is
\begin{equation}
 \begin{split}
 S_N
 =\frac{\ii}{2}\int\dd\tau\,\dd\sigma\sqrt\gamma\,
 \bm\psi^T
 \Bigl[&
 D_\tau
 +\Aone[\bm V]
 \\
 &+\bm M
 -\Athree[\bm\eta_3]
 \Bigr]\bm\psi.
 \end{split}
 \label{eq:multi_majorana_action}
\end{equation}
Here \(\bm V\) and \(\bm\eta_3\) are real symmetric matrices, while
\(\bm M\) is real and antisymmetric.  The matrix operators are defined by
\[
\Aone[\bm V]=\bm V\partial_s+\tfrac12\partial_s\bm V,
\qquad
\Athree[\bm\eta_3]
=\tfrac12(\bm\eta_3\partial_s^3+\partial_s^3\bm\eta_3),
\]
with derivatives acting componentwise.  Geometry can enter every matrix
through \(K\), \(v_n\), and their derivatives.  A local orthogonal rotation of
the Majorana basis transforms all three matrices and generates temporal and
spatial flavor connections.  It can simplify the connection locally, but it
does not remove \(\bm M\) independently of \(\bm V\) and
\(\bm\eta_3\); global holonomy and invariant spectral data remain.

For co-propagating modes, an antisymmetric constant $\bm M$ mixes the
Majoranas and shifts their zero crossings but cannot gap the net chiral
central charge.  Counterpropagating pairs can be gapped by allowed mass terms.
The surviving low-energy central charge is fixed by bulk anomaly matching,
whereas the detailed number of reconstructed branches and their velocities is
not.

For a multichannel neutral sector, the anomaly-controlled net chiral part of the stress
response is obtained by replacing
\begin{equation}
 c_\psi=\frac12
 \quad\longrightarrow\quad
 c_{\rm neutral}=\frac12(N_R-N_L)
 \label{eq:central_charge_generalization}
\end{equation}
in the anomaly-controlled net chiral part.  The full response generally
contains separate poles or continua from inequivalent velocities and from
nonchiral pairs.

The Moore--Read and Halperin 331 states have equal electrical Hall
conductance.  Electron tunneling and strong interactions can localize the
charge modes at their interface, leaving one propagating chiral Majorana
mode~\cite{Yang2017}.  The single-field geometric action developed above then
applies directly to the neutral mode.  Because $\Delta\nu=0$, however, the
charged swept-area mechanism of Eq.~\eqref{eq:bosonic_shape_bracket} is absent.
A freely moving Moore--Read--331 domain wall requires a separate shape
symplectic structure, potentially from geometric response, microscopic
interfacial dipole dynamics, or coupling to other gapless modes
~\cite{ReadRezayi2011,ParkHaldane2014}.  The neutral
Majorana theory alone does not determine the motion of the wall.

Pfaffian--anti-Pfaffian interfaces at $\nu=5/2$ support several possible
boundary phases with different patterns of charge gapping and neutral-mode
hybridization~\cite{MrossOregStern2018,Hsin2020}.  Numerical work
finds strong mode reconstruction and an intrinsic electric dipole tied to the
mismatch of guiding-center Hall viscosity~\cite{ParkHaldane2014,ZhuShengYang2020}.  A single Majorana coupled to a shape field is therefore not a universal
low-energy description of this interface.  The appropriate starting point is Eq.~\eqref{eq:multi_majorana_action}
combined with the remaining bosonic modes and with geometric-response terms
that encode the dipole and Hall-viscosity mismatch.

The worldsheet formulation nevertheless provides a useful organization:
spatial relabeling acts on all modes, extrinsic geometry enters through local
matrix coefficients, and the net gravitational anomaly fixes the ungappable
chiral central charge.  Determining the actual low-energy matrix
$\bm V(K,v_n)$ for a Pfaffian--anti-Pfaffian domain wall requires microscopic
matching, not topology alone.

The single-channel description assumes that the neutral sector has not
reconstructed into additional co- and counterpropagating Majoranas.  Recent
microscopic work proposes precisely such a neutral reconstruction near
realistic $\nu=5/2$ edges~\cite{ZhangWuHutasoitJain2014,Lotric2025}.  In that regime the
no-backscattering result of Sec.~\ref{sec:propagation} applies only after all
counterpropagating pairs have been integrated out.  Before that reduction,
geometry and disorder can mix the branches and produce localization or
mass-domain-wall zero modes.  The one-channel results apply only after such additional branches have been
gapped or integrated out.

\subsection{Matching and diagnostics}
\label{sec:matching}

The worldsheet coefficients divide naturally into universal and microscopic
data.  Topology fixes the net chirality, \(c_\psi=1/2\) for a single
right-moving Majorana, and the allowed spin structures.  Geometry fixes the
measure, the convective derivative, the half-density transport term, and the
kinematic vertices generated by a moving line element.  The coefficients
\(v_\psi\), \(\eta_3\), \(\lambda_K\), \(\lambda_n\), and their multichannel
generalizations must instead be obtained from a microscopic interface model.

The straight-interface spectrum determines $v_\psi$ and $\eta_3$ from
Eq.~\eqref{eq:majorana_dispersion}.  Circular droplets of several radii then
separate the $R^{-2}$ curvature shift $\lambda_K$ from the $R^{-3}$
coefficients in Eq.~\eqref{eq:circle_spectrum_expanded}.  The fit must retain
the NS/R sector and the global parity projection.  The bosonic Moore--Read
parent Hamiltonian used in Refs.~\cite{LiMa2021,TurkerYang2022} is a natural
benchmark because the bulk state is exact and the bosonic and fermionic edge
branches are readily distinguished.

For a prescribed nonuniform static interface, the Majorana level spacing or
propagation time determines the time-of-flight functional.  The
first-derivative theory predicts
\begin{equation}
 \mathcal T[K]
 =\oint\frac{\dd s}
 {v_\psi+\lambda_KK+\lambda_{K^2}K^2+\cdots}.
 \label{eq:matching_time_of_flight}
\end{equation}
The absence of reflection tests the validity of the single-chiral-channel
description.  Observed backscattering would diagnose extra neutral
branches, a velocity zero, or significant higher-gradient physics.

A weak oscillatory deformation at momentum $k$ should produce neutral
spectral weight concentrated near $\omega=v_\psi k$ with the low-momentum
scaling
\begin{equation}
 \mathcal S_{K}(\omega,k)
 \propto
 c_\psi\lambda_K^2k^7
 \delta(\omega-v_\psi k)
 \label{eq:curvature_absorption_scaling}
\end{equation}
in the ideal linear theory.  On a circle, the geometric form factor in
Eq.~\eqref{eq:circle_curvature_form_factor} multiplies the CFT norm in
Eq.~\eqref{eq:Virasoro_transition_norm}; ratios between topological sectors at
fixed $n$ are independent of that common factor.  The same quantities are accessible to exact diagonalization, matrix-product-state calculations,
or real-time simulation of an effective edge Hamiltonian.

The central-charge-controlled nonlocal pole contributes, when expanded on
the long-wavelength bosonic branch, to the \(k^7\) shape dispersion,
Eq.~\eqref{eq:self_energy_low_k}.  Local neutral-sector renormalizations are
absorbed into the geometric coefficients and need not begin only at this
order.  A calculation that extracts both the static six-derivative shape energy and the frequency-dependent Majorana response can separate a bare geometric coefficient from
the neutral dynamic correction.  The absence of small-$k$ damping provides an independent check.

\section{Conclusion}
\label{sec:discussion}

To conclude, we have developed a spatially reparametrization-invariant theory for a
chiral Majorana mode carried by a freely moving quantum Hall interface.
The changing line element fixes the transport and normalization of the
fermion: in scalar variables it produces the half-density term in the
equation of motion, while in canonically normalized variables it gives the
standard equal-time anticommutator and the action of spatial relabeling.
This kinematic structure is distinct from the additional curvature- and
velocity-dependent couplings, whose coefficients depend on microscopic
interface physics.

The construction also clarifies the physical difference between static and
dynamical geometry.  Within the leading one-derivative, single-channel
theory, a smooth static velocity profile changes only the accumulated time of
flight and cannot backscatter or localize a lone chiral Majorana as long as
the local velocity remains nonzero.  A time-dependent deformation instead
couples to the Majorana stress tensor and allows energy exchange with the
shape sector.  The nonlocal part of this response is fixed by the Ising
central charge, although its overall coupling and local contact terms are
nonuniversal.  On a closed droplet, the same stress algebra produces
topological-sector-dependent transition weights, subject to the charged-sector
and fermion-parity constraints of the full Moore--Read theory.

These results provide the neutral counterpart of the charged worldsheet
kinematics developed in the companion paper~\cite{Ma2026I}. Together, the two theories
separate the universal ingredients inherited from Hall response and Ising
topological order from the geometric coefficients that must be determined
microscopically.  The single-channel description applies when additional
neutral branches are absent or have already been gapped.  It does not provide
the shape dynamics of an interface with $\Delta\nu=0$, and it is not a
universal description of reconstructed Pfaffian--anti-Pfaffian or
$\nu=5/2$ edges.

Several problems remain.  The coefficients controlling the Majorana velocity,
dispersion, and geometric couplings should be extracted from microscopic
droplet and interface calculations.  A complete quantum treatment must also
incorporate the bulk gravitational response, global spin structure, charged
sector, and fermion-parity projection.  Multichannel interfaces require a
simultaneous treatment of neutral reconstruction, disorder, and interfacial
dipole response.  Open interfaces and junctions introduce genuine Majorana
boundary conditions and fusion-space dynamics and form the natural next step
in the effective string theory of quantum Hall interfaces.

\begin{acknowledgments}
The author is sincerely grateful to Professor Kun Yang for introducing him to
quantum Hall interface physics, and for the insightful discussions and
encouragement during his stay at the National High Magnetic Field Laboratory.
Although the author has since followed a path outside academia, the present
work continues those discussions.
\end{acknowledgments}

\appendix
\section{Variation of the Majorana action}
\label{app:variation}

Consider
\begin{equation}
 S=\frac{\ii}{2}\int\dd\tau\,\dd\sigma\sqrt\gamma\,\psi A\psi.
 \label{eq:appendix_general_action}
\end{equation}
Here \(A^\dagger\) denotes the formal adjoint with respect to the spacetime
measure \(\dd\tau\,\dd\sigma\sqrt\gamma\), up to temporal endpoint terms.  For
Grassmann-odd $\psi$, integration by parts gives
\begin{equation}
 \delta S
 =\frac{\ii}{2}
 \int\dd\tau\,\dd\sigma\,\sqrt\gamma\,
 \delta\psi(A-A^\dagger)\psi.
 \label{eq:appendix_variation_formula}
\end{equation}
For $D_\tau=\partial_\tau-\beta\partial_\sigma$, integration by parts gives
\begin{equation}
 \begin{split}
 \int\dd\tau\,\dd\sigma\sqrt\gamma\,fD_\tau g
 =-\int\dd\tau\,\dd\sigma\sqrt\gamma\,
 \Bigl[D_\tau f
 &+\bigl(\partial_\tau\ln\sqrt\gamma
 \\
 &-\gamma^{-1/2}\partial_\sigma(\sqrt\gamma\beta)\bigr)f
 \Bigr]g.
 \end{split}
\end{equation}
Using
\begin{equation}
 \partial_\tau\ln\sqrt\gamma
 =\partial_sv_t-Kv_n,
 \qquad
 \gamma^{-1/2}\partial_\sigma(\sqrt\gamma\beta)
 =\partial_sv_t,
\end{equation}
one obtains the formal adjoint
\begin{equation}
 D_\tau^\dagger=-D_\tau+Kv_n.
\end{equation}
Hence the antisymmetric part is
\begin{equation}
 \frac12(D_\tau-D_\tau^\dagger)
 =D_\tau-\frac12Kv_n.
\end{equation}
Similarly,
\begin{equation}
 (f\partial_s)^\dagger
 =-f\partial_s-f_s,
\end{equation}
which proves Eq.~\eqref{eq:A1_definition}.  Since
$(f\partial_s^3)^\dagger=-\partial_s^3f$, antisymmetrization gives
Eq.~\eqref{eq:A3_definition}.

\section{Canonical graded bracket and relabeling generator}
\label{app:canonical}

For the half-density kinetic term
\begin{equation}
 L_{\rm kin}=\frac{\ii}{2}\int\dd\sigma\,\chi\dot\chi,
\end{equation}
the momentum constraint is
\begin{equation}
 \zeta(\sigma)
 =\pi_\chi(\sigma)-\frac{\ii}{2}\chi(\sigma)\approx0.
\end{equation}
Using left Grassmann functional derivatives and the graded canonical bracket
\(\{\chi(\sigma),\pi_\chi(\sigma')\}
=\delta(\sigma-\sigma')\), the constraint
matrix is proportional to $-\ii\delta$, and its inversion gives
Eq.~\eqref{eq:chi_Dirac_bracket}.

Varying
\begin{equation}
 G_\psi[\xi]
 =\frac{\ii}{2}\int\dd\sigma\,\xi\chi\chi'
\end{equation}
with respect to $\chi$ gives
\begin{equation}
 \frac{\delta G_\psi}{\delta\chi}
 =\frac{\ii}{2}
 \left(2\xi\chi'+\xi'\chi\right).
\end{equation}
Eq.~\eqref{eq:chi_Dirac_bracket} therefore yields
\begin{equation}
 \{\chi,G_\psi[\xi]\}_D
 =\xi\chi'+\frac12\xi'\chi.
\end{equation}
A second application gives the classical Witt algebra in
Eq.~\eqref{eq:fermion_Witt}.  Normal ordering of the quantum stress tensor
produces the central term in Eq.~\eqref{eq:majorana_Virasoro}.

\section{Reference conventions for the circular Majorana}
\label{app:circle}

For a circle of length $L=2\pi R$, write
\begin{equation}
 \psi(s)
 =\frac{1}{\sqrt L}
 \sum_{r\in\mathbb Z+\alpha}
 \psi_r e^{\ii rs/R},
 \qquad
 \{\psi_r,\psi_{r'}\}_+=\delta_{r+r',0}.
\end{equation}
The spin-structure parameter is $\alpha=1/2$ in the NS sector and $\alpha=0$
in the R sector.  The quadratic Hamiltonian corresponding to
Eq.~\eqref{eq:majorana_covariant_action} is
\begin{equation}
 H_\psi
 =\frac12\sum_r
 \left(
 \frac{V_R}{R}r+\frac{\eta_3}{R^3}r^3
 \right):\psi_{-r}\psi_r:,
\end{equation}
which gives Eq.~\eqref{eq:circle_single_particle_spectrum}.  The normal-ordering
constant of the linear conformal term is $-c_\psi/24$ on the cylinder; the
allowed primary weights are listed in Eq.~\eqref{eq:Ising_weights}.

\section{Monge-gauge expansion}
\label{app:Monge}

In static gauge,
\begin{equation}
 \sqrt\gamma D_\tau
 =\sqrt{1+u_x^2}\,\partial_t
 -\frac{u_x\dot u}{\sqrt{1+u_x^2}}\partial_x.
\end{equation}
Moreover,
\begin{equation}
 \sqrt\gamma\,\partial_s=\partial_x.
\end{equation}
These identities prove Eq.~\eqref{eq:exact_Monge_kinematic_action}.  Expanding
\begin{align}
 \sqrt{1+u_x^2}
 &=1+\frac12u_x^2+O(u^4),
 \\
 \frac{u_x\dot u}{\sqrt{1+u_x^2}}
 &=u_x\dot u+O(u^4)
\end{align}
gives Eq.~\eqref{eq:kinematic_quadratic_vertices}.  The geometric scalars have
linear expansions
\begin{align}
 K&=u_{xx}+O(u^3),
 \\
 v_n&=\dot u+O(u^3),
 \\
 \partial_sK&=u_{xxx}+O(u^3).
\end{align}
These expansions give Eq.~\eqref{eq:linear_shape_majorana_vertex}.  In the stress-tensor
form of the interaction, the velocity coupling contributes
\begin{equation}
 \Delta L_{\lambda_n}=-\lambda_n\int\dd x\,\dot u\,T_\psi,
\end{equation}
and therefore shifts the momentum conjugate to the shape according to
\begin{equation}
 P_u=P_u^{(0)}-\lambda_nT_\psi+\cdots.
\end{equation}
This is the sense in which the \(v_nT_\psi\) operator is a mixed kinetic
coupling rather than an ordinary potential term.

\section{Stress correlator and Virasoro transition weight}
\label{app:stress}

At zero expectation value of $T_\psi$, the central part of the equal-time
commutator is
\begin{equation}
 \langle[T_\psi(x),T_\psi(0)]\rangle
 =\frac{\ii c_\psi}{24\pi}\partial_x^3\delta(x).
\end{equation}
Chiral propagation implies
\begin{equation}
 T_\psi(t,x)=T_\psi(0,x-v_\psi t).
\end{equation}
Therefore
\begin{equation}
 \begin{split}
 \chi^R_{TT}(\omega,k)
 ={}&-\ii\int_0^\infty\dd t\int\dd x\,
 e^{\ii\omega t-\ii kx}
 \\
 &\times\frac{\ii c_\psi}{24\pi}
 \partial_x^3\delta(x-v_\psi t),
 \end{split}
\end{equation}
which evaluates to Eq.~\eqref{eq:TT_response_full}, up to local contact terms.

For the Fourier convention \(u(t,x)\propto e^{\ii(kx-\omega t)}\), the
general linear interaction in Eq.~\eqref{eq:linear_vertex_stress_form} has
source
\begin{equation}
 g(\omega,k)
 =\lambda_Kk^2+\ii\lambda_n\omega+\ii\lambda_{sK}k^3.
\end{equation}
With the convention
\(D_R^{-1}=D_{0,R}^{-1}+\Pi_u^R\), the induced retarded kernel is
\begin{equation}
 \Pi_u^R(\omega,k)
 =g(-\omega,-k)g(\omega,k)\chi^R_{TT}(\omega,k),
\end{equation}
which gives Eqs.~\eqref{eq:shape_self_energy_general} and
\eqref{eq:general_stress_source}.

For a primary state $|h\rangle$ satisfying $L_n|h\rangle=0$ for $n>0$,
\begin{equation}
 \begin{split}
 \langle h|L_nL_{-n}|h\rangle
 &=\langle h|[L_n,L_{-n}]|h\rangle
 \\
 &=2nh+\frac{c}{12}(n^3-n).
 \end{split}
\end{equation}
This proves Eq.~\eqref{eq:Virasoro_transition_norm}.  At \(n=2\) and
\(c_\psi=1/2\), the three neutral Ising modules give
\begin{equation}
 \mathcal W_{1,2}=\frac14,
 \qquad
 \mathcal W_{\psi,2}=\frac94,
 \qquad
 \mathcal W_{\sigma,2}=\frac12,
\end{equation}
and hence the ratio in Eq.~\eqref{eq:n2_Ising_weight_ratio}.


\begin{thebibliography}{99}

\bibitem{MooreRead1991}
G. Moore and N. Read,
Nonabelions in the fractional quantum Hall effect,
Nucl. Phys. B \textbf{360}, 362 (1991).

\bibitem{MilovanovicRead1996}
M. Milovanovi\'c and N. Read,
Edge excitations of paired fractional quantum Hall states,
Phys. Rev. B \textbf{53}, 13559 (1996).

\bibitem{ReadGreen2000}
N. Read and D. Green,
Paired states of fermions in two dimensions with breaking of parity and
time-reversal symmetries and the fractional quantum Hall effect,
Phys. Rev. B \textbf{61}, 10267 (2000).

\bibitem{LiMa2021}
Q. Li, K. K. W. Ma, R. Wang, Z.-X. Hu, H. Wang, and K. Yang,
Dynamics of quantum Hall interfaces,
Phys. Rev. B \textbf{104}, 125303 (2021).

\bibitem{TurkerYang2022}
O. T\"urker and K. Yang,
String-like theory of quantum Hall interfaces,
Phys. Rev. B \textbf{106}, 245138 (2022).

\bibitem{Ma2026I}
K. K. W. Ma,
An effective string theory toolbox for quantum Hall interfaces I:
Worldsheet kinematics and constraint structure,
companion manuscript (2026).

\bibitem{Yang2017}
K. Yang,
Interface and phase transition between Moore--Read and Halperin 331
fractional quantum Hall states: Realization of chiral Majorana fermion,
Phys. Rev. B \textbf{96}, 241305(R) (2017).

\bibitem{LevinHalperinRosenow2007}
M. Levin, B. I. Halperin, and B. Rosenow,
Particle-hole symmetry and the Pfaffian state,
Phys. Rev. Lett. \textbf{99}, 236806 (2007).

\bibitem{LeeRyuNayakFisher2007}
S.-S. Lee, S. Ryu, C. Nayak, and M. P. A. Fisher,
Particle-hole symmetry and the $\nu=5/2$ quantum Hall state,
Phys. Rev. Lett. \textbf{99}, 236807 (2007).

\bibitem{MrossOregStern2018}
D. F. Mross, Y. Oreg, A. Stern, G. Margalit, and M. Heiblum,
Theory of disorder-induced half-integer thermal Hall conductance,
Phys. Rev. Lett. \textbf{121}, 026801 (2018).

\bibitem{ZhuShengYang2020}
W. Zhu, D. N. Sheng, and K. Yang,
Topological interface between Pfaffian and anti-Pfaffian order in the
$\nu=5/2$ quantum Hall effect,
Phys. Rev. Lett. \textbf{125}, 146802 (2020).

\bibitem{Hsin2020}
P.-S. Hsin, Y.-H. Lin, N. M. Paquette, and J. Wang,
An effective field theory for fractional quantum Hall systems near
$\nu=5/2$,
Phys. Rev. Research \textbf{2}, 043242 (2020).

\bibitem{ZhangWuHutasoitJain2014}
Y. Zhang, Y.-H. Wu, J. A. Hutasoit, and J. K. Jain,
Theoretical investigation of edge reconstruction in the $\nu=5/2$ and $7/3$
fractional quantum Hall states,
Phys. Rev. B \textbf{90}, 165104 (2014).

\bibitem{Lotric2025}
T. Lotri\v{c}, T. Wang, M. P. Zaletel, S. H. Simon, and
S. A. Parameswaran,
Majorana edge reconstruction and the $\nu=5/2$ non-Abelian thermal Hall
puzzle,
arXiv:2507.07161.

\bibitem{WanYangRezayi2006}
X. Wan, K. Yang, and E. H. Rezayi,
Edge excitations and non-Abelian statistics in the Moore--Read state: A
numerical study in the presence of Coulomb interaction and edge confinement,
Phys. Rev. Lett. \textbf{97}, 256804 (2006).

\bibitem{HuRezayiWanYang2009}
Z.-X. Hu, E. H. Rezayi, X. Wan, and K. Yang,
Edge-mode velocities and thermal coherence of quantum Hall interferometers,
Phys. Rev. B \textbf{80}, 235330 (2009).

\bibitem{ReadRezayi2011}
N. Read and E. H. Rezayi,
Hall viscosity, orbital spin, and geometry: Paired superfluids and quantum
Hall systems,
Phys. Rev. B \textbf{84}, 085316 (2011).

\bibitem{KaneFisher1997}
C. L. Kane and M. P. A. Fisher,
Quantized thermal transport in the fractional quantum Hall effect,
Phys. Rev. B \textbf{55}, 15832 (1997).

\bibitem{CappelliHuertaZemba2002}
A. Cappelli, M. Huerta, and G. R. Zemba,
Thermal transport in chiral conformal theories and hierarchical quantum Hall
states,
Nucl. Phys. B \textbf{636}, 568 (2002).

\bibitem{AlvarezGaumeWitten1984}
L. Alvarez-Gaum\'e and E. Witten,
Gravitational anomalies,
Nucl. Phys. B \textbf{234}, 269 (1984).

\bibitem{Stone2012}
M. Stone,
Gravitational anomalies and thermal Hall effect in topological insulators,
Phys. Rev. B \textbf{85}, 184503 (2012).

\bibitem{BradlynRead2015}
B. Bradlyn and N. Read,
Topological central charge from Berry curvature: Gravitational anomalies in
trial wave functions for topological phases,
Phys. Rev. B \textbf{91}, 165306 (2015).

\bibitem{GromovJensenAbanov2016}
A. Gromov, K. Jensen, and A. G. Abanov,
Boundary effective action for quantum Hall states,
Phys. Rev. Lett. \textbf{116}, 126802 (2016).

\bibitem{SinhaBradlyn2025}
S. Sinha and B. Bradlyn,
Extrinsic geometry and gappable edges in rotationally invariant topological
phases,
Phys. Rev. B \textbf{112}, 155306 (2025).

\bibitem{StoneRoy2004}
M. Stone and R. Roy,
Edge modes, edge currents, and gauge invariance in $p_x+\ii p_y$
superfluids and superconductors,
Phys. Rev. B \textbf{69}, 184511 (2004).

\bibitem{MorozHoyosRadzihovsky2016}
S. Moroz, C. Hoyos, and L. Radzihovsky,
Chiral $p\mathbin{\pm}\ii p$ superfluid on a sphere,
Phys. Rev. B \textbf{93}, 024521 (2016).

\bibitem{Quelle2016}
A. Quelle, C. Morais Smith, T. Kvorning, and T. H. Hansson,
Edge Majoranas on locally flat surfaces: The cone and the M\"obius band,
Phys. Rev. B \textbf{94}, 125137 (2016).

\bibitem{NayakReview2008}
C. Nayak, S. H. Simon, A. Stern, M. Freedman, and S. Das Sarma,
Non-Abelian anyons and topological quantum computation,
Rev. Mod. Phys. \textbf{80}, 1083 (2008).

\bibitem{DiFrancesco1997}
P. Di Francesco, P. Mathieu, and D. S\'en\'echal,
\emph{Conformal Field Theory}
(Springer, New York, 1997).

\bibitem{Ginsparg1988}
P. Ginsparg,
Applied conformal field theory,
in \emph{Fields, Strings and Critical Phenomena},
Les Houches, Session XLIX, edited by E. Br\'ezin and J. Zinn-Justin
(North-Holland, Amsterdam, 1990), arXiv:hep-th/9108028.

\bibitem{FendleyFisherNayak2007}
P. Fendley, M. P. A. Fisher, and C. Nayak,
Edge states and tunneling of non-Abelian quasiparticles in the
$\nu=5/2$ quantum Hall state and $p+\ii p$ superconductors,
Phys. Rev. B \textbf{75}, 045317 (2007).

\bibitem{ParkHaldane2014}
Y. Park and F. D. M. Haldane,
Guiding-center Hall viscosity and intrinsic dipole moment along edges of
incompressible fractional quantum Hall fluids,
Phys. Rev. B \textbf{90}, 045123 (2014).

\end{thebibliography}
\end{document}